\documentclass[12pt]{article}
\usepackage{amsmath,amsfonts,amscd} 
\usepackage{graphicx}
\usepackage{enumerate}
\def\bra{\langle} 
\def\ket{\rangle} 
 
\def\C{\mathbb{C}} 

\def\N{\mathbb{N}} 

\def\R{\mathbb{R}}
\def\Z{\mathbb{Z}}
\def\Zp{{\Z_p}}
\def\etc{{ etc.\ }} 
 
\def\ie{{ i.e.\ }} 
\def\hc{\dagger}
\def\RV{\texttt{Rig Veda\ }}
\begin{document}
\title{Consciousness and quantum mechanics of macroscopic systems}
\author{Mikhail V. Altaisky \\  
\small\em Laboratory of Radiation Biology, 
Joint Institute for Nuclear Research, \\
\small\em Joliot-Curie 6, Dubna, 141980, Russia; \\
\small\em and \\
\small\em Space Research Institute RAS, 
Profsoyuznaya 84/32, Moscow, 117997, Russia\\
\small\em E-mail: altaisky@mx.iki.rssi.ru}
\date{April 20, 2008}
\maketitle

\begin{abstract}
We propose the quantum mechanical description of complex systems should be 
performed using two types of causality relation: the ordering relation ($x\prec y$) and the subset 
relation ($A\subseteq B$). The structures with two ordering 
operations, called the {\em causal sites}, have been already proposed in context of quantum gravity (Christensen and Crane, 2005). 
We suggest they are also common to biological physics and may describe how the brain works. 
In the spirit of the Penrose ideas we identify the geometry 
of the spacetime with universal field of consciousness. The latter has its evident 
counterparts in ancient Indian philosophy and provides a 
framework for unification of physical and mental phenomena. 
\end{abstract}

Key words: consciousness, quantum measurement, spacetime

\section{Introduction}
The creation of quantum mechanics at the beginning of the previous century was,
perhaps, the most significant event when the discovery of particular physical 
effects resulted in global change of human understanding of the world. The 
radical point of view suggested by quantum mechanics consists in complete 
disclaim of the objective reality: for the state of object -- as declared by 
quantum mechanics -- depends on what the observer knows. Later, the widely 
accepted statistical interpretation of quantum mechanics  
pacified the situation by assertion that quantum mechanical probability 
 describes an ensemble of quantum particles rather than a single 
one, and so it statistically predicts the share of particles to be found in a given state 
in certain experiment. Using this paradigm one can surely do a lot of 
quantum mechanical calculations with a good agreement with experiment 
without paying serious attention whether the world exists or not. 

In our days the attitude to the problem of consciousness and measurement in 
quantum mechanics has changed, 
basically due to the studies related to quantum computations \cite{DJ1992,Keyl2002}. It becomes 
utmost evident, or at least very attractive, to assume that final observation  on any 
quantum system is performed by consciousness -- the ultimate observer. At the same time, 
the research in brain science and the attempts to simulate complex brain wave dynamics by 
dynamical systems, in particular by neural networks \cite{RA2006}, reveal a deep parallelism between the 
hypothetical procedure of quantum computations and information processing in human 
brain \cite{Ingber1983,Penrose1989,BE1992}. 
Alternatively it seems utmost impossible to explain tremendous amount of information 
processing per second, performed by human brain at room temperature, without significant 
thermodynamic heat losses, those should be of order $\Delta Q = k T \Delta I \ln 2$, that is 
about $3\cdot 10^{-21}$ J/operation at $T=300^oK$.
Qualitatively the parallels between quantum phenomena in micro-world and psychological 
phenomena in brain science are quite well known. For instance, asking a person what he is thinking about 
this very moment, will cause a process similar to wave function collapse: from all thoughts superposed 
in his mind before being asked he will select only one thought and formalise it in his answer -- 
that is an einselection process \cite{Orlov1982}. 

The importance of consciousness in quantum physics was understood much earlier than the question of 
brain information processing arose. In quantum physics the result of measurement can not be described 
independently of the observer, \ie cannot be described ``objectively'' \cite{mes,Mensky2000}.
If a quantum system has two possible states, $|\psi_1\ket$ and $|\psi_2\ket$, their linear superposition 
\begin{equation}
|\psi\ket = c_1|\psi_1\ket + c_2 |\psi_2\ket
\label{sup1}
\end{equation}
is also a possible state of the same system. 
The states of quantum system, defined as vectors in abstract Hilbert space, are determined only 
relatively to certain {\em process of measurement}: 
if there is a Hermitian operator $A^\hc = A$, such that 
\begin{equation}
A |\psi_1\ket = a_1 |\psi_1\ket, \quad A |\psi_2\ket = a_2 |\psi_2\ket, \quad a_1\ne a_2 \in \R,
\end{equation}
the states $|\psi_1\ket$ and $|\psi_2\ket$ are referred to as eigenstates of the physical {\em observable} $A$. 
Spin, momentum, energy, coordinate, \etc are physical observables. 

The process of measurement, \ie determination of the state of quantum system leads to the collapse of 
the linear superposition of states \eqref{sup1} to the measured state -- either of possible ones. 
The probability of certain result of measurement is  given by squared complex amplitude:
\begin{equation}
|\psi\ket = c_1|\psi_1\ket + c_2 |\psi_2\ket \to 
\begin{cases} 
|\psi_1\ket, & p_1 = |c_1|^2 \\
|\psi_2\ket, & p_2 = |c_2|^2, 
\end{cases}
\label{red1}
\end{equation}
where the initial state is normalised $|c_1|^2+|c_2|^2=1$. 
The equation \eqref{red1} is the simplest form of the von Neuman reduction postulate, that states the 
collapse of the wave function \eqref{sup1} from a linear superposition of possible states to a definite 
state. 

The process of measurement takes place by means of interaction of the 
system with the environment, but is not caused by that interaction only. Indeed,
if the measuring  device was in a state $|\phi\ket$ before the measurement, and works so, that depending 
upon the detected state of the measured system $|\psi\ket$, the device undergoes either of quantum transitions, 
$$
\begin{cases}
|\phi\ket \to |\phi_1\ket, & \hbox{if the system was in the state } |\psi_1\ket, \\
|\phi\ket \to |\phi_2\ket, & \hbox{if the system was in the state } |\psi_2\ket,
\end{cases}
$$
then the result of measurement is that the state of the combined system (``system $+$ device'') undergoes 
a transition from factorisable to non-factorisable superposition state:
\begin{equation}
|\psi\ket |\phi\ket \to c_1 |\psi_1\ket |\phi_1\ket + c_2 |\psi_2\ket |\phi_2\ket.
\label{entl}
\end{equation}
The latter is called an {\em entangled} state, emphasising the fact, that the states of the system and 
that of the measuring device are no longer mutually independent after the measurement. The transformation 
\eqref{entl} itself does not lead to the collapse of wave function, since it retains the superposition of 
states. The collapse happens at the moment when the states of the the measuring device are {\em discriminated 
by observer}; that means that different states of the measuring device should be mutually orthogonal 
$$\bra \phi_1 | \phi_2 \ket = 0;$$
if it is not the case the linear superposition will survive after the measurement.  

The entanglement arising in course of measuring introduces two essential problems into quantum measurement: 
({\em i}) the problem of nonlocality, and ({\em ii}) the problem who is the observer. The former makes the 
events separated by space-like intervals causally connected: measuring the projection of spin of one EPR 
particle we know the projection of its counterpart \cite{EPR1935}. The latter drives the final observer, 
who discriminates the states of measuring device, out of physical reality: a state of the measuring device 
$|\phi\ket$, which is entangled with the state of the studied quantum system $|\psi\ket$, is measured by the 
other device $|\phi_1\ket$, which becomes entangled with $|\phi\ket$ and so on, including human vision system, 
parts of brain \etc, -- also described by quantum mechanics, -- but all these systems themselves cannot {\em discriminate}.
So, it should be something beyond this to say ``this and not this'' and record the result. This something we define 
as an ultimate observer, the function of which is the {\em awareness}.   

Therefore before the observer became aware of the result of measurement he describes the system by wave 
function \eqref{sup1}, but when he became aware of its state he starts to use either with 
$|\psi_1\ket |\phi_1\ket$ 
or $|\psi_2\ket |\phi_2\ket$. This means the result of measurement is dependent upon whether or not the observer 
sees the measuring device.

To summarise these strange facts of quantum reality, from very beginning of quantum mechanics, it was suggested that 
the wave function of a 
quantum system may evolve in two essentially different ways: (1) exhibit linear unitary evolution, governed by 
the Schr\"odinger equation; (2) exhibit quantum collapse when being observed; (3) the wave function 
in quantum mechanics describes not a quantum system {\em per se}, but our perception of that quantum 
system. This three points 
comprise the Copenhagen interpretation of quantum mechanics, developed by N.Bohr, V.Heisenberg and W.Pauli.
 According to Copenhagen interpretation \cite{ATDN}, the task of science is to extend the range of our experience and to 
predict the results of our sensations after certain actions. The quantum mechanical amplitudes describe 
therefore the probabilities of our sensations. For this very reason the Copenhagen interpretation has 
been attacked many times as denying the existence of the external world.  
The description of our sensations instead of material world is not 
bad itself: it simply means there should be something beyond our sensations, 
which is not less important than the matter, and which should be included 
in our picture of the world. It can be referred to as universal consciousness.

In  present paper we tried to consider the matter and the consciousness 
as equal constituents of the world, trace their roots in cosmology, and 
compare the present view on consciousness and the geometry of the 
spacetime with ancient Vedic philosophy. The remainder of this paper 
is organised as follows. In {\em section \ref{hier:sec}} we consider 
the relation of consciousness to cosmology and topology in view of ancient Indian philosophy.
It is claimed that a hierarchic type of causality (causal sites) should be used instead 
of point causality. 
In {\em section \ref{brain:sec}} we discuss  our perception of the world as macroscopic 
quantum effect. An hierarchic Hilbert space formalism is proposed in  
{\em section \ref{inf:sec}} to generalise the idea of hierarchic information states, 
proposed to describe the brain dynamics in terms of dynamical systems. A few general 
remarks on consciousness and causality in living and non-living matter are presented 
in {\em Conclusion}.

\section{Hierarchical systems: How the geometry emerges \label{hier:sec}}
The positive solution of the problem of measurement may be given by understanding of common source of matter and 
consciousness. Such ideas have been proposed since long ago, at least 
since Pythagoras, but mostly at the level of metaphysics. Rather 
recently a participation of gravity in the wave function collapse in brain has been proposed by R.Penrose 
\cite{Penrose1989}. This idea is very attractive mathematically: both the gravity and the consciousness 
impose certain sets of relations on matter, and it is natural to suggest these two types of relations 
have common source. 
To illucidate that common source let us borrow the picture of the origin of the 
world drawn in ancient scriptures, in particular in Indian philosophy. 
As it is said in Rig Veda \cite[page 13]{RV}:
\begin{verse} 
There was neither being nor non-being,\\ 
Without breath breathed by its own power That One\\
\emph{RV.10.129, Creation hymn} 
\end{verse} 
In Sanskrit language ``That One'' is denoted by the term {\sl brahman}.
Non-differentiable {\sl Brahman} is the source of differentiated (\ie  
comprised of more than one entity) world. 
In {\sl Chandogya upanishad (Chand 4.10.5)} \cite[page 16]{RV} we read 
``{\sl brahman} is void''. In modern terms this can be understood as vacuum.

According to \RV the creation of the world -- ``creation of many from 
One'', ``manifested from non-manifested'' -- results in three fundamental 
entities
\begin{description}
\item[Prakriti] -- the term close to our word ``matter''
\item[Atman] -- there is no exact English equivalent to it -- it is often 
translated as ``universal soul'', ``objective soul'', or ``pure consciousness''
\cite[page 27]{RV}
\item[Purusha] - eng. ``Person'' -- that, who realise its identity to {\sl atman}
(``I am He'') -- the term related to human consciousness.
\end{description}

The hymns of \RV are written in an ambiguous manner, and their unique interpretation is quite 
difficult \cite{Muller,RV-ling}. Here we cite the \RV text basically for illustration -- the same question on 
the origin of the present Universe from something one and unique could be asked in purely 
physical settings: How present continuous space has emerged? The Big Bang theory does not 
answer this question for it is based on continuous spacetime itself. To get a continuous 
space from something discrete it is required to have ({\em i}) sufficiently big discrete set of 
objects ({\em ii}) with sufficiently big set of relations on that set. Typical example is the 
spatial grid -- a set of vertexes (balls) with binary neighbouring relations between them; 
depending on the number of neighbours we have different dimensions of the space.

However the neighbouring relations are not the only possible relations. The other type 
of possible relations is {\em the part -- the whole} relation. It is related to the fragmentation 
of a single object into a number of parts (subsets). Such relations are described by hierarchic 
trees. In view of this we can assume the existence of universal set of relations between 
objects ({\sl prakriti}) understanding this set as a universal consciousness ({\sl atman}). 

The term {\sl purusha}, mentioned above, is more complex. By definition, {\sl purusha} is ``that 
who realises its identity to {\sl atman}''; it is the source of ego, free will, etc., and is 
attributed to the field of psychology, rather than physics. It is important, however, that 
{\sl atman} -- the objective consciousness -- is the {\em universal} set of relations. Therefore 
the {\sl purusha} should be subset of it. This means in the final end of any observation the 
final observer is the universal consciousness -- {\sl atman}. (Some authors considering the matter-consciousness dualism in view of Indian philosophy use the 
{\sl prakriti-purusha} dihotomy, restricting the consciousness to observer's consciousness.)

Assuming the {\sl atman} to be a universal set of relations imposed on matter  let us consider a 
mathematical toy-model of the origin of continuous world from a non-discriminated {\sl brahman}  
($|0\ket$), which we understood as vacuum. 

The formation of matter ({\sl prakriti}) could be described by action of creation operators to 
vacuum state $|i\ket=a_i^+ |0\ket, \quad i=\overline{1,N}$. This results in a tree-like process 
which yields a discrete, but not necessarily countable set of excitation
\begin{equation}
\begin{array}{cccccc}
         &                 & |1\ket & \to                & |11\ket & \cdots \\
         &\nearrow  & \vdots & \searrow    & \vdots  &        \\
 |0\ket  &                 & \vdots &                    & |1M\ket &        \\
         &\searrow  & \vdots &                    &         &        \\
         &                 & |N\ket & \ldots             &         &
\end{array}
\label{tlp}
\end{equation}
The tree-like process \eqref{tlp} can be continued or not with either of the first generated excitations 
$\{ |i\ket \}_{1,N}$.
We assume it is the way the {\sl prakriti} emerged from {\sl brahman}, but not the geometry. It is important, that 
in physical sense the {\sl prakriti} is understood as pre-matter, rather than matter. The matter fields 
of observed elementary particles can be combined of pre-matter elements: this requires 
certain relations to combine those elements into a set -- the geometry is required. 
These relations are to be imposed by {\sl atman}. 
What kind of relation those could be?
Definitely the 
relations that distinguish between ``this and not this'' and binary relations 
between elements. The most trivial topology will be the space of all possible 
subsets of the set of {\sl prakriti} elements, with each subset being understood 
as an open set. Interestingly, the ``atoms'' introduced by Democritus philosophy 
were derived from the non-differentiable absolute by imposing the relation that 
distinguish between ``this and not this''.
However, this construction deals with a discrete set of objects, and we need 
something more to explain how the continuous geometry emerges from a discrete 
set, and how this set is ordered to explain the observed causal phenomena. 

\begin{figure}
\centering \includegraphics[width=6cm]{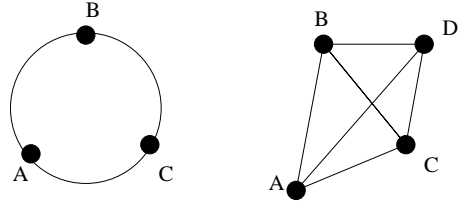}
\caption{Constructing $S^1$ and $S^2$ spheres from sets of points}
\label{s12:pic}
\end{figure}

There is a question how the discrete set of relations can result in a 
continuous differentiable manifold. The first way to answer, suggested by 
combinatorial topology, is that a set of binary relations between points 
(elements of {\sl prakriti}) can be considered as a complex, with points being the 
vertices. For instance, a set of 3 points $\{A,B,C\}$ with the set of 
binary relations $\{\chi(A,B),\chi(B,C),\chi(C,A)\}$ is equivalent to 
the $S^1$ sphere; similarly, the set of 4 points $\{A,B,C,D\}$ with 
corresponding binary relations  
$$\{\chi(A,B),\chi(B,C),\chi(C,A),\chi(A,D),\chi(B,D),\chi(C,D)\},$$
\ie a three-dimensional simplex, is 
equivalent to the sphere $S^2$, see Fig.~\ref{s12:pic}.
The metrics on this space can be introduces as $\rho(x,y)=$ the shortest path 
from vertex $x$ to vertex $y$ of the graph, with the distance between 
{\em neighbouring}, \ie those subjected to the relation $\chi(\cdot,\cdot)$, vertexes 
considered as unity. 
To introduce a chronological-like ordering a more complicated set of relations 
is required. First, the partial ordering like that described in 
\cite{Sorkin2003} or \cite{CC2005} can be introduced by regarding the 
formation of the Universe topology a {\em multiplicative process} of diversification \eqref{tlp}. The idea of multiplicative process as the 
origin of the Universe with fractal Cantorian geometry has been 
discussed by many authors 
\cite{Ord1983,AS1995,Nottale1996,Elnaschie1998,Sidharth2001}. However, as to 
my knowledge, it was not considered as a source of causality relations.

\begin{figure}
\centering \includegraphics[width=6cm]{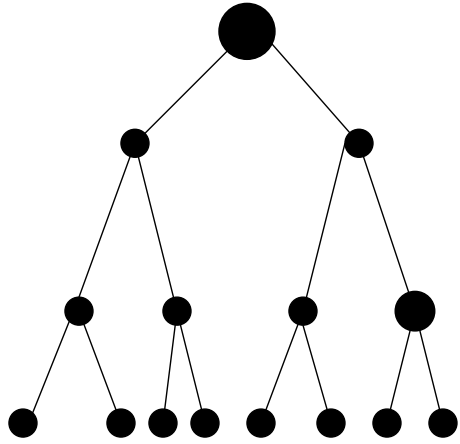}
\caption{Binary multiplicative process}
\label{tree:pic}
\end{figure}
We assume that the vertical ordering (shown in Fig.~\ref{tree:pic}) corresponds to $\subset$  
and the horizontal ordering corresponds to $\prec$ of \cite{CC2005}. 
In quantum description of pre-spacetime the evolution can be observed 
by changing the number of entities and the quanta of time is just an action of 
the next creation operator on the already existed level 
$a_i^+(t+1) a_j^+(t)\ldots a_k^+(1)|0\ket$. This quanta of time are presumably 
equal to $\tau_{Pl}\sim 10^{-42}$sec. 
If there are no changes - there is no time. The observer of course observe not 
the pre-atoms, or regions, shown in Fig.~\ref{tree:pic}, but some groups of them 
together with certain relations -- all this is perceived as {\em fields}.
The horizontal ordering $\prec$ corresponds to the signal propagation. 
Where we understand a signal can be generated by any change of physical 
fields.

On a discrete set generated by a multiplicative process \eqref{tlp} from a single 
parent object we have two kinds of relations:

\begin{enumerate}[a]

\item The neighbouring relations make this space metrisable: 

if $\chi(a,b)=true\ \& \ a\ne b$ then $\rho(a,b)=1$.
The distance between points which are not direct neighbours is counted as a shortest path composed 
of neighbouring pairs:
\begin{equation}
a, x_1,x_2,\ldots,x_n, b: \chi(a,x_1)=\chi(x_1,x_2)=\ldots \chi(x_n,b) =  true.
\label{npath}
\end{equation}  
The neighbouring relation $\chi(\cdot,\cdot)$ implies the equivalence relation $a\sim b\ \& \ b\sim c \Rightarrow a\sim c$, 
where $a\sim b$ means the existence of a path \eqref{npath} between $a$ and $b$. Therefore the 
neighbouring relation defines the partition of the initial set into disjoint classes.
If the neighbouring relation is equipped with a partial order 
\begin{equation}
a \prec b\ \& \ b \prec c \Rightarrow a\prec c,
\label{porder}
\end{equation}
with closed loops avoided, then the manifold constructed from the neighbouring relations will inherit 
that partial order. Physically, the partial order corresponds to the time ordering. 
A signal can propagate from point $a$ to point $b$ only if there exists a causal path 
\begin{equation}
a, x_1,x_2,\ldots,x_n, b: a \prec x_1 \prec x_2 \prec \ldots \prec x_n \prec b. 
\label{cpath}
\end{equation}

\item The inheritance (parent-child) relation is another type of causal relations, not related to 
the causal paths of the type ``a''. Inheritance is a relation between parent and child objects, the 
whole and the part. A typical physical example is the EPR pair: two fermions, being produced by a 
single (parent) boson retain quantum correlations between their states, regardless the fact that they 
are not connected by causal path of type ``a''. The inheritance from parent to child $(C\subseteq P)$ also 
implies a partial order
\begin{equation}
A \subseteq B\ \& \ B \subseteq C \Rightarrow A\subseteq C.
\end{equation}
\end{enumerate}

The causal relations of type ``a'' are purely classical -- they are common to differential geometry, Newtonian 
mechanics and general relativity. The type ``b'' causal relations can not be found in classical physics, in which 
the coordinates can be measured with arbitrary high precision. The partial order $\subseteq$ describes 
the refinement of the measurement: one can measure (affect) a part of something only having affected the whole first; 
but not reverse. In EPR experiments the states of two different fermions are measured only after 
the state of the parent boson have been affected. In this way the multiplicative process \eqref{tlp} with two 
ordering relations, $\prec$ and $\subseteq$, unifies the geometry and the theory of measurement. 
 The question is how these two relations are imposed on the discrete set generated by a multiplicative 
process.   
An interesting point of view is considering the Universe in terms 
of information theory, as big processor which includes all, where the matter 
fields are computational elements immersed in computational media -- consciousness \cite{Zuse1967}.

\section{Consciousness and brain \label{brain:sec}}
Turning to biological science we face the problem of relations between brain and consciousness. If the universal consciousness ({\sl atman}) is as fundamental as matter is, 
if the geometry of the world exists by virtue of it, what is its relation to 
the personal observer's consciousness, or mind?
Mind (sanscr. {\sl chitta}), viz. a human mind, is a category of much less generality 
than {\sl atman}, and is subjunctive 
to the latter. Mind  is the ``software'' that works in brain. 
Its task is to process the sensory inputs, to form response, and to control its execution. The dynamics of 
neuron firings, as emphasised by Tegmark \cite{Tegmark2000}, can be described by classical or semiclassical 
approximation, however the potential and the forces related to the activity of 
neurons are governed by the 
field of consciousness, in the sense, that neurodynamics works in highly unequilibrial regime \cite{RA2006}, and it is the 
consciousness that drives the neuron system to either of its possible attractors. Thus the trajectory of an 
atom in a living cell can be evaluated only at time intervals much less than the typical times of the 
consciousness dynamics.

The difference between living and non-living systems lies in the fact, that living systems have something, 
perhaps what was called {\sl purusha} in ancient Indian philosophy, that makes them different from just a 
mechanical collection of parts. It is said in \RV scriptures, that each living being has its separate, but identical 
{\sl purusha}.  For non-living matter, which does not have it, -- say for a stone, the product of wave functions of all atoms 
in that stone completely determines its quantum state of that stone.  For a living object, in contrast, 
the state of the whole defines the state of the parts, but not vice versa. That is why the entropy of a living 
system does not increase according to the second law of thermodynamics. The individual consciousness, {\sl purusha}, 
is the field that determines the dynamics of a living system and decreases entropy. 
It was even suggested by a physicist N.Umov a hundred years ago to introduce the third law of thermodynamics specially for it to meet Darwin's evolutionary ideas.
 
The ``{\em mind}'', in contrast to consciousness, is something more special to psychology and
can be attributed only to leaving creatures, if not only for humans. The problem of mind and consciousness have
been studied by philosophers from ancient times, and it seems be helpful to recall certain functions of mind
to understand how the brain works \cite{Amit1989,Stapp1993}. According to ancient Indian philosophy the mind ({\sl chitta})
has at least three 
functional components: {\sl manas} -- the recording facility, {\sl buddhi} -- the intellect, and {\sl ahamkara} -- the ego
\cite{patanjali}. This three functions enables the data acquisition about the state of the environment and 
the state of body, develop possible plans of action and control their execution. 
All those provide the 
survival of the organism. For instance, the {\sl manas} registers (by sense organs) a fast moving object, the {\sl buddhi} 
evaluates 
its size and possible trajectory and finds out it will soon approach my personal position and transmits 
the information to {\sl ahamkara}. If it is the case the {\sl ahamkara} commands the body to change position in an 
appropriate way to avoid the collision. 

If the work of the {\sl manas} - the recording facility - can be explained physically, at least in principle, in a 
way similar to the operation of digital camera, the operation of {\sl buddhi} is more complex. Here we use 
Sanskrit word {\sl buddhi} instead of English ``intelect'', for very often the intellect is understood is purely 
algorithmic decision taking, but {\sl buddhi} includes both algorithmic and non-algorithmic evaluation of input 
information. The former is described by formal rules and can be implemented on computer, the latter cannot be 
implemented in such a way and it is the place for quantum aspects of consciousness. 

To understand how the brain works different models have been proposed, see 
\cite{Arbib1987,Eccles1991} for review. Generally the brain is considered 
as an information processing unit that maps certain input information into certain output information. 
However, the word ``information'' requires definition itself. First, according to \cite{NP1977} the amount of 
information is a number of dichotomy questions (to be answered yes/no) we have to ask to describe the system completely. 
Therefore the definition of {\em information} implicitly requires {\em consciousness} -- the reader 
of the  yes/no answers. 
Even having known the state of each neuron in the 
brain, we do not have a unique way to determine what thoughts are going on. The situation is similar to 
traditional computing: having measured the currents in microprocessor we still can 
not say what software is 
running. 
The consciousness in a wide sense, {\sl atman}, as was already mentioned, is out of material world by definition:
it is a set of relations imposed upon material objects. In a narrow sense, the consciousness, or more exactly the mind,
keeps the scheme of the organism (body), makes the scheme of the external world and maintains the body-world 
interactions. Being a part of the universal consciousness ({\sl atman}), the mind ({\sl chitta}) is also capable of 
observing its own state.  
 
The real thinking process includes along with the verbal information space, which is 
processed classically, certain images, associations, \etc, that can not be formalised using 
a finite and universal alphabet \cite{Fauconnier1994}. By analogy with quantum computations, we can consider such space 
as a Hilbert space of images. This seems to agree with the Copenhagen interpretation 
of quantum mechanics, that says the wave function describes our perceptions, rather than 
reality itself \cite{Bohm1951}. This does not contradict the statement, that between two ``measurements'' the evolution 
of mind can be fairly well described by a dynamical system. In the space of all states 
of consciousness such dynamical system forms a trajectory -- the stream of 
consciousness \cite{Khrennikov2002}.

It is not rather clear, at what lowest neurophysiological level the quantum state reduction 
correlates with consciousness. One of the candidates is the level of cytoskeleton \cite{HHT2002}. 
Penrose also suggested the so-called single-graviton criterion, that means the quantum gravity effects may
be essential at the level of the single neuron \cite{Penrose1989}. However the
final detector of any quantum measurement is the observer's consciousness and therefore the quantum 
collapse happens at the level of macroscopic brain activity \cite{Stapp1993}. 

The usual role of the observer's consciousness ({\sl chitta}) is registering the state of a macroscopic 
device, a {\em pointer}, -- say a light pointer, or a digital indicator. The latter (possibly via a long
chain of intermediate mesoscopic devices) interacts with the measured quantum system. The 
whole chain should exhibit linear evolution according to the Schr\"odinger equation. 
The von Neuman collapse happens because the states of the 
pointer device are utmost mutually orthogonal  for the tremendous number of degrees of freedom 
involved
$$ 
\bra i|j \ket \sim e^{-N},$$ 
where $N$ is the number of degrees of freedom in the device. Because of this orthogonality the reduction 
\eqref{red1} takes place. 
 
Let us imagine the quantum system interacting with very few degrees of freedom of human observer. 
Such effects do really take place: a human eye in dark-adapted state is capable of detecting a single 
photon \cite{Stryer1985,Sakmar1998}. 
A photon registered {\em in vivo} by human observer may be the cause of entanglement between a quantum system 
and remote observer. Indeed, if a photon is emitted in Raman scattering, the observer's consciousness 
becomes entangled with the 
molecule which scattered the photon. To some extent, this means the observer becomes aware of the state of that molecule. 
The important point of such consideration is that the measurement process, performed by human consciousness, consists of two stages \cite{Orlov1982}. The last is the quantum transition, but the first is {\em taking attention} of the system to be studied. 
The quantum transition is described by the probability amplitudes, while taking attention is a voluntary act, and is 
not described by the probability amplitudes -- it just imposes certain relations upon neurons. This means, that 
if a person concentrates on an external system, and wish 
to observe it, he makes his consciousness entangled with that system. 

The difference from a standard observation in visual range, -- when the observer receives a lot of photons with 
uncorrelated phases from the studied object, -- is that a volition to observe a system makes the observer's consciousness and 
the system into a combined system, but does not necessarily lead to a collapse. The {\em collapse} happens when the 
observer {\em records} his impressions in verbal form -- thoughts, -- 
doing so he destroys the coherent superposition and 
makes it into a single information state of his consciousness, which corresponds to a definite pattern of neural activity. 
 
If the consciousness and the topology of spacetime have the same origin -- both are sets of relations imposed on matter, -- 
the measurement performed by direct perception, \ie by brain being entangled with the studied system, can also change the state of observed system. The possible mechanism 
is the same as for the EPR-pair: having measured the projection of spin of one particle of the pair we, therefore, fix the 
projection of its counterpart on another end. Such hypothesis is implicitly supported by the fact, that most significant effects 
of intended change of state by paranormal inductors (psi-phenomena) were observed for semiconductor devices, where 
the number of half-spin excitation over forbidden zone is easily controlled \cite{Dulnev2000}. 

\section{Hierarchic structure of information states \label{inf:sec}}
\begin{quotation}
`` ... however the complex object may be, the thought of it is one individual 
state of consciousness.'' \\
 William James 
\end{quotation} 
The term {\em state of consciousness} in the above phrase requires clarification. If the object is really 
complex, \ie,  it contains a number of hierarchic levels, the knowledge of that object is acquired by 
taking snapshots of different hierarchic levels by means of sequential quantum measurements. 
That is why 
it is natural to assume the state of consciousness, which is a representation of that object, to be 
a hierarchic tree. This coincides with the definition of information state as a $p$-adic tree given 
in \cite{Khrennikov2002}.
If the information is coded in a string written in $m$-letter alphabet, possibly 
a chain of neurons with $m$ states \cite{Eccles1974}, and 
$$x=(x_0,x_1,\ldots,x_i,\ldots), \quad x_i=\overline{0,m-1}, \quad m\in \N,$$ is an 
information vector, where the coordinate $x_0$ is the most important, $x_1$ 
is less important, and so on, then different information states can be labelled
by $m$-adic trees, or $m$-adic integers $x=x_0+x_1 m + x_2 m^2+\ldots$. 

\begin{figure}
\centering \includegraphics[width=6cm]{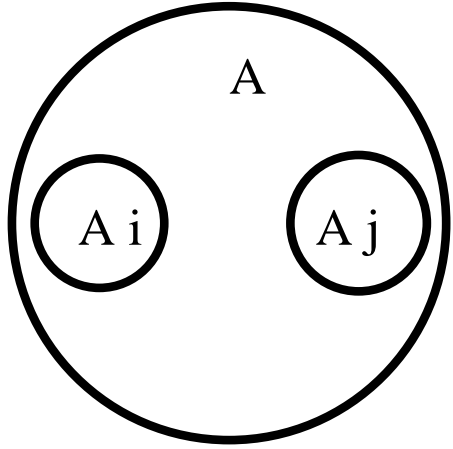}
\caption{Hierarchic system $A$ with two subparts $i$ and $j$, that carry the index of the whole, 
since their state is constrained by the state of the whole}
\label{hc:pic}
\end{figure}

In quantum mechanics, if we have a system $A$ consisting of two parts, -- say, a 
meson consisting of quarks and anti-quark, -- the wave function of the whole 
$A$ is completely determined by direct product of the constituents wave 
functions:
$$
\Psi_M = \bar q \otimes q.
$$
Such system is unconscious: there is nothing extra in meson, that can not 
be found in the product of its components wave functions. However, it is 
formally possible to represent the state of meson in the hierarchic form \cite{Alt03IJQI}
$$
\left\{ |A\ket,  
\{ |Ai\ket, |Aj\ket \}
\right\},
$$
see Fig.\ref{hc:pic}.
The possible spin states of meson are $|\uparrow\ket, |0\ket, |\downarrow\ket$; the 
spin states of quarks are $|\uparrow\ket, |\downarrow\ket$. 
If there are two states of meson, say $|A\ket = |\uparrow\ket,\quad |B\ket=|\downarrow\ket$ their linear superposition  
\begin{equation}
\Psi=\alpha \left\{ |A\ket, \{ |Ai\ket,  |Aj\ket\}  \right\} +
\beta  \left\{ |B\ket, \{ |Bi'\ket, |Bj'\ket\} \right\} 
\label{ls}
\end{equation}  
is also allowed state. 
Perhaps, for a nonliving matter, such as meson, the index $A$ in \eqref{ls} is 
redundant -- for the state of the whole is just a product of the states of its 
constituents. For biological systems the presence of the state of 
the whole as an {\em independent argument} is important. To some extent the same 
is required for any open system. The wave function of an electron in molecule is 
not the same as the wave function of a free electron. One can 
argue that formally we can take into account all other electrons and nuclei 
degrees of freedom of that molecule and calculate the wave function of the 
whole molecule, but this is practically impossible: the interaction with 
environment and, finally, the wave function of the whole Universe should be 
taken into account. So, for hierarchic systems, such as atoms, molecules and 
bigger complexes, it is quite reasonable to treat the state of the whole as 
independent argument. 
Strictly speaking, the measurement procedure itself demands to keep the 
state of the whole as separate independent argument when studying  the 
behaviour of the parts. Say, the measurement of the spin state of quark in 
meson can be performed only after the whole meson was prepared in a certain 
quantum state. 

In biological settings, the state of the whole is controlled 
by consciousness and cannot be expressed as a product of its constituents 
states -- so it must be kept independently. Reproducing the classical 
consideration of \cite{Khrennikov2002}, we can make the same statement about thinking 
process: we can retrieve details of a thought only when it was first grasped 
as a whole, and then we go down to the parts (details). Each step on this thought-tree 
corresponds to a measurement procedure.   

Generally the thoughts of material objects have hierarchic structure. 
If before the measurement took place the studied object was in 
a superposition of quantum states 
\begin{equation}
|\phi\ket = \alpha_1 |[x^{(1)}]\ket +\alpha_2 |[x^{(2)}]\ket +\ldots \alpha_n |[x^{(n)}]\ket,
\end{equation}
where $|[x^{i}]\ket$ correspond to different information states,
then after the measurement only one of the alternatives will survive, 
let it be the $i$-th state, survived
with probability $|\alpha_i|^2$: it will form a classical information state defined 
in hierarchic tree. This coincides with the definition of information state as a $p$-adic tree. 
If $[x^{(i)}]$ is a classical information state, \ie the string of integers 
$x^{(i)}_0, x^{(i)}_1,x^{(1)}_2,\ldots, x^{(i)}_k<p$, the corresponding $p$-adic tree 
can be labelled by a $p$-adic integer 
\begin{equation}
x^{(i)} = x^{(i)}_0 + x^{(i)}_1 p + x^{(i)}_2 p^2 + \ldots.
\label{ptree}
\end{equation}     
The vector in hierarchic Hilbert space \cite{Alt03IJQI}, which corresponds to $p$-adic tree 
\eqref{ptree}, will be written as 
\begin{equation}
|[x]\ket = \bigl\{ p^0|x_0\ket, p^{-1/2}|x_0 x_1\ket, p^{-1}|x_0 x_1 x_2\ket,
\ldots  \bigr\}
\label{hivec},
\end{equation}
with scalar product, norm \etc, being defined component-wise:
\begin{equation}
\bra [x] | [y] \ket = \bra x_0 | y_0 \ket + p^{-1}\bra x_0 x_1 | y_0 y_1 \ket + p^{-2}\bra x_0 x_1 x_2| y_0 y_1 y_2\ket +
\ldots.
\label{scprod}
\end{equation}
The representation \eqref{hivec} of $p$-adic trees states for the oriented 
string of characters: $x_0$ is more important than $x_1$, and so on. 
If instead one would try to use 
$$
|x\ket = \bigl\{ |x_0\ket, |x_1\ket, |x_2\ket,  \bigr\} 
$$
the orientation will be lost. 

\begin{figure}
\centering \includegraphics[width=6cm]{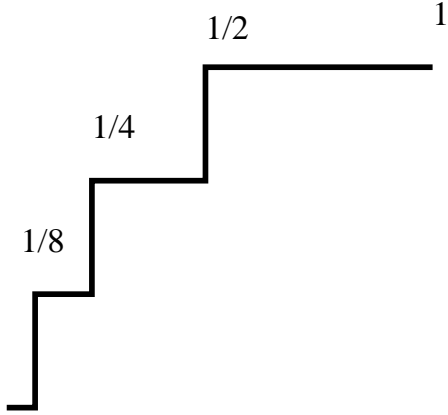}
\caption{Spatial domains corresponding to piecewise constant behaviour of wave-function 
in $p$-adic norm, $p=2$.}
\label{hs:pic}
\end{figure}

The link to the quantum mechanics of $c$-valued wavefunctions of $p$-adic argument 
normalised as 
\begin{equation}
\phi: \Zp \to \C : \|\phi\|^2 = \int_{\Zp} |\phi(x)|^2 dx < \infty
\label{zpnorm} 
\end{equation} 
might be given by setting the scale behaviour of the hierarchic wave function 
to 
\begin{equation}
\phi(x) \sim |x|_p^{1/2} \le 1, x \in \Zp.
\end{equation}
This corresponds to geometric interpretation: smaller $p$-adic norm means smaller spatial scales. 
On the larger scale the domain of piece-wise constant behaviour is larger, see Fig.\ref{hs:pic}.

In our hierarchic formalism the function $\phi(x)$ is understood in terms of hierarchic states 
$$
\phi(x) \equiv \bra [\phi] | [x] \ket,$$
where  $[x]$ and $[\phi]$ belong to the space \eqref{hivec} and the space conjugated to it, 
respectively. 
The measurements on such a system should be described by a tree of operators, each of them corresponding 
to its particular scale -- from the most rough to the finest:
\begin{equation}
\hat A \equiv   |x_k\ket A_0^{kl} \bra x_l| +  \frac{1}{p}|x_k x_m\ket A_1^{klmn} \bra x_l x_n| + \ldots
\end{equation}

We ought to say that hierarchic representation of external 
world in brain in the form of hierarchic trees -- information states -- stems from the discriminative 
feature of consciousness: discrimination between ``this'' and ``not this'' leads to a tree-like 
structure. In any observation (measurement) we start from rough and continue with subtle details.
There is no account for this fact in standard quantum mechanics. According to Copenhagen interpretation 
quantum mechanics predicts {\em our sensations} after certain actions. We infer that 
our representation of the external world is correct if our predictions coincide with observations.
An important property of consciousness is its ability ``to be aware of itself''. This means 
consciousness can observe (measure) itself. Thus the dynamic of thinking, and the brain activity, 
constrained by that dynamics, can be described by quantum mechanics of information states represented 
by vectors in hierarchic Hilbert space \eqref{hivec}. Such dynamics be relevant to the brain thinking 
processes on intervals between measurements. In the same way as the geometry of spacetime defines 
the laws of motion for unconscious matter, the consciousness defines the laws of motion for living 
matter. It makes relations between subparts of living systems in such a way that the living matter 
acts against the second law of thermodynamics.  

\section{Conclusion}
In traditional approach to quantum mechanics, declared by N.Bohr, life is considered as something 
complementary to the procedure of quantum measurement. If we measure position of an atom in living cell with 
a high accuracy, the energy, required to achieve this accuracy, will kill the cell. According to 
this paradigm, the physics study interactions between objects, which are separable, \ie, may be discriminated 
from each other. Biology, in contrast, studies the effect of changing the state of the whole (organism, organ, cell) 
to the functioning of its parts. Those two methods have been considered as complementary to each other. 
We suggest, that this seeming complementarity stems from the fundamental structure of the world imposed by universal 
consciousness. 

The separate existence of the mind and the matter has been proclaimed even by Descartes. The new finding, made by R.Penrose, 
was that consciousness, which is responsible for the measurement, and the geometry, which is responsible for 
the constancy of laws of nature, may be just the same.

In our paper, basing on ancient Indian philosophy along with modern physics, we claim there are two types 
of consciousness: 
objective consciousness, which can be identified deterministic laws of physics, and subjective 
consciousness, which is responsible for the behaviour of living beings.
The geometry of spacetime that we observe is formed by those two types of consciousness, by imposing 
two types of causal relations, preceding ($\prec$) and inclusion ($\subseteq$). The former is more known 
in physics, the latter -- in biology; both, in fact, constitute a general causal structure, known as causal 
site.
Our individual consciousness, which process the information on the external world and is also capable of observing 
itself, represents the results of observations in hierarchic trees, or information states. This hierarchy 
is comes from the general structure of the objective consciousness -- the structure of causal site. Mathematical 
description are hierarchic quantum states corresponding to information states.     

\section*{Acknowledgements}
The author is thankful for hospitality to V\"axj\"o University (Sweden), where the draft of this paper 
have been prepared, and to Drs. A.Khrennikov and B.G.Sidharth for useful comments. The support of the DFG project 436 RUS 113/951 is acknowledged.

\end{document}